# Cybersecurity Issues and Practices in a Cloud Context: A Comparison Amongst Micro, Small and Medium Enterprises.

## Full research paper

### Ruwan Nagahawatta
School of Accounting, Information Systems and Supply Chain
RMIT University
Victoria, Australia.
Email: S3863329@student.rmit.edu.au

### Sachithra Lokuge
School of Business
University of Southern Queensland
Queensland, Australia.
Email: ksplokuge@gmail.com

### Matthew Warren
Centre for Cyber Security Research and Innovation
RMIT University
Victoria, Australia
University of Johannesburg, South Africa
Email: matthew.warren2@rmit.edu.au

### Scott Salzman
Department of Information Systems and Business Analytics
Deakin University
Victoria, Australia.
Email: scott.salzman@deakin.edu.au

## Abstract

The advancement and the proliferation of information systems among enterprises have given rise to understanding cybersecurity. Cybersecurity practices provide a set of techniques and procedures to protect the systems, networks, programs and data from attack, damage, or unauthorised access. Such cybersecurity practices vary and are applied differently to different types of enterprises. The purpose of this research is to compare the critical cybersecurity threats and practices in the cloud context among micro, small, and medium enterprises. By conducting a survey among 289 micro, small and medium-sized enterprises in Australia, this study highlights the significant differences in their cloud security practices. It also concludes that future studies that focus on cybersecurity issues and practices in the context of cloud computing should pay attention to these differences.

# 1  Introduction

The landscape of cloud computing has significantly changed over the last decade with more cloud service providers offering numerous services, and cloud infrastructures (Goode et al. 2015; Pahl et al. 2017; Walther et al. 2018). However, the advancement of such technologies creates various issues and challenges for enterprises (Dolawattha et al. 2020; Kaluarachchi et al. 2020; Sedera and Lokuge 2019; Varghese and Buyya 2018). For example, issues such as security, governance, cost management, and resource management are some of the prominent ones. In the traditional cloud, there are significant security risks related to data storage and hosting of multiple users, mitigated by robust mechanisms to guarantee user and user data isolation (Chejerla and Madria 2017; Nagahawatta et al. 2021a). The advancement of fog computing ecosystem further creates more issues as the risks are of more significant concern since a wide range of nodes are accessible to users (Gupta et al. 2013; Varghese and Buyya 2018). In addition, selection of the right cloud products and services may also present a challenge, and organisations are presented with too many choices (Pahl et al. 2017).

The problem is that addressing trust, privacy, and security issues in advanced technology remains a challenge because it needs a combination of technological and non-technical approaches that often lie outside the governance of an organisation (Kaluarachchi et al. 2020; Lowry et al. 2017; Moody et al. 2018). Privacy is one of the core issues in the use of information technology (IT), including the need for policy components during integration, protect identity information and transaction histories (Khayer et al. 2020). Customers may be concerned about information stored in the cloud being accessed by others anywhere in the world (Nagahawatta et al. 2020; Ratten 2014). Cloud security refers to the controls, services and policies that protect cloud applications, infrastructure, and data from threats (Zissis and Lekkas 2012). Cyber-attacks can be caused due to negligence and ignorance which is intolerable, making this a primary area to be reviewed (Nagahawatta et al. 2021b).

However, application of such practices and their challenges may vary depending on the size of the enterprise. For example, the nature of the resources, the extent of enterprise and the type of the business determines the severity of the challenges and the nature of the practices. This paper focuses on the cybersecurity practices and issues faced by small and medium sized enterprises (SMEs). SMEs play a significant part in the economic progress of any nation in the world (Wong and Aspinwall 2004), and they have attracted increasing and strong attention from the Australian government. Australian SMEs comprise 95% of businesses and over 57% of gross domestic product (GDP) contribution. Large Australian enterprises adopt more than 70% of commercial payment cloud computing services; however, the adoption of SMEs is less than 40% due to various challenges (Australian Bureau of Statistics, 2019). Recent studies in Australia on SMEs cloud computing adoption indicate that security challenges are one of the critical factors that discourages SMEs to adopt cloud computing (e.g., Asiaei and Rahim 2019; Senarathna et al. 2018).

While analysing the SME segment, it is evident that prior research considers them as one segment with common characteristics. Interestingly, in recent years, research highlights the need to analyse them separately and SMEs were further analysed under micro, small and medium sized enterprises. It is always a sensitive issue to make general claims about SMEs, because their diversity may be their most important characteristic (Belás et al. 2015). For example, among the micro, small, and medium enterprises, medium enterprises can technically distinguish themselves better than micro and small firms (Gupta and Barua 2016). As such, a new area of research focus has been established on diversity within the SME segment (Belás et al. 2015). These categories are classified in terms of the number of employees (Gupta et al. 2015), capital investments, annual turnover (Belás et al. 2015; Gibson and Vaart 2008). As a result of these distinguishing characteristics, it seems reasonable to assume that cybersecurity issues and practices may vary by SME enterprise size. However, a limited literature in this area has been devoted to studying the cybersecurity issues and practices of these different categories. Therefore, this study will address this research gap by assessing the differences in cybersecurity practices and issues among micro, small, and medium enterprises. To analyse this, in this study we employed a quantitative approach to examine cybersecurity issues and practices in a cloud context among Australian micro, small and medium enterprises.

The paper is structured as follows. The following section will provide the background of the study by providing related work in this domain. The third section provides the methodology used for the data collection and analysis. Then, the fourth section provides the overview of the results, and a discussion is provided in the next section. Finally, the paper summarises its contribution and recognises future research areas.

## 2 Related Work

While cloud computing provides significant business benefits and opportunities, it also presents security, privacy, and trust related challenges, particularly for SMEs (Ayong and Naidoo 2019; Khayer et al. 2020; Nagahawatta and Warren 2020; Salim et al. 2015). A survey undertaken by the Australian Cyber Security Centre (ACSC) among Australian SMEs show an increasing number and extent of cyber-attacks targeting SMEs. This further highlights the importance of understanding cybersecurity practices among different categories of Australian SMEs (ACSC 2020).

### 2.1 Characteristics of SME Categories

There is no generally accepted definition for SMEs, and the existing definitions vary from country to country, from sector to sector, over time, even among institutions within a single country. Generally, the definition of SMEs entails the criteria such as the value of capital, amount of revenue, number of employees, estimate of fixed assets, and the nature of the organisation or combination of those factors. The Australian Bureau of Statistics (ABS) defines SMEs as enterprises with 1 to 199 employees. There are three types of SMEs: (i) micro businesses which have fewer than 4 employees, (ii) small businesses with 5 to 19 employees and (iii) medium businesses with 20 to 199 employees (ABS 2019). Several previous studies indicated that size of the enterprise (i.e., micro, small and medium enterprises) is considered as one of the fundamental variables in explaining enterprise behaviour (Heidt et al. 2019; Redondo and Fierro 2007). Paik (2011) indicated that company size is a leading factor for enterprise performance. The enterprise size is a factor commonly used to set boundaries between segments of enterprises (Tornatzky and Fleischer 1990). Cataldo, Pino and McQueen (2020) suggested that variation in IT adoption exists among micro, small, and medium enterprises.

As discussed, SMEs generally vary from large enterprises in terms of their capacity, structure, and business size (Sedera 2016). For SMEs, cost is a crucial issue as they have limited financial resources to spend on their IT facilities. In general, SMEs possess problems such as lack of IT resources, lack of IT professionals and security systems (Doherty 2015; Khayer et al. 2020). However, SMEs have closer communication between employees and management, and the ability to execute and implement decisions quickly. The key challenge that SMEs face is keeping their costs under control and hence, many cannot allocate much of their budget to IT. Anecdotal studies show that there exist some important similarities and differences between micro, small and medium enterprises regarding the cybersecurity issues and practices associated with the cloud environment.

A study by Rahayu and Day (2015) found that medium enterprises are more willing to adopt the Internet for business use compared to smaller and micro sized enterprises. Yeboah-Boateng and Essandoh (2014) found that enterprise size has a significantly positive relationship with cloud computing adoption. SMEs tend to have a small management team and centralised management style. The manager of an SME is usually the business owner and has a strong influence on the enterprise's decision making (Heidt et al. 2019). Amini and Bakri (2015) state that SMEs' IT infrastructure is not well established compared to that of larger enterprises. Furthermore, they have limited resources and often inconsistent revenue (Heidt et al. 2019; Mijnhardt et al. 2016). Amini and Bakri (2015) suggest that despite these differences, different types of SMEs have common features such as inadequate planning, unsophisticated software, or IT applications that tended to have greater risks and uncertainty compared with large organisations. Further, SMEs perceive those characteristics of security and privacy in the cloud to be more complicated and not as easy to identify when compared to large organisations.

### 2.2 Cybersecurity Threats and Vulnerabilities

The scope of cybersecurity issues extends to the security of IT systems deployed in enterprises as well as to the broader digital networks including critical national infrastructures (Ani 2017). Unfortunately, preliminary security surveys by government such as ACSC (2020) show an increasing number of cyber threats targeting enterprises, but with a lack of information about the characteristics of the attacks and their possible impacts. Therefore, it is important to analyse existing cybersecurity threats, vulnerabilities, and their solutions with a comprehensive view of the cybersecurity, to gain a complete picture of the cybersecurity practices in relation to SMEs. New technologies offer vital business opportunities and benefits. However, it provides, privacy, security, and risk issues, particularly for SMEs (Mijnhardt et al. 2016).

The cybersecurity attacks and threats are diverse in terms of motivation and technological exploits. Ranging from insider attacks motivated by malice to the accidental misconfiguration of enterprise networks, from a lack of contingency planning to automated exploitation of known security vulnerabilities (Sinha et al. 2019). The loss of user control can be problematic in situations such as data damage or misuse, unauthorised access, unavailability, or infrastructure failure (Ani 2017). Due to the lack of awareness SMEs can face several vulnerabilities. For instance, a lack of competence can make any enterprise vulnerable to social engineering attacks like phishing. If an employee is not competent enough to differentiate a legitimate business email from a phishing email, it puts an organisation at risk (Symantec 2019). Furthermore, the lack of concern for professional development and failing to update personal knowledge, can lead to a lack of awareness of current cyber threats such as DDoS (Distributed Denial of Service) attacks, social engineering, hacktivism, and credential harvesting malware (ACSC 2020). This reveals that employees do not practice due diligence, which could lead to a multitude of financial and legal risks.

## 2.3 Cybersecurity and Australian SMEs

According to the ACSC, cybercrime is a threat to Australia's national and economic prosperity, with cybercrime expertise improving and tradecraft being adapted to target specific businesses. Cybercrime will continue to be an attractive option for criminals, as it enables them to generate large profits with a low risk of identification and interdiction (ACSC 2020). The ACSC (2020) report highlighted that Australian SMEs know cybersecurity is important regardless of how they rate their understanding of cybersecurity. However, they face significant barriers when attempting to implement good cybersecurity practices in particularly for micro firms. These barriers include limited budgets, an absence of IT staff for support, the complex field of cybersecurity, challenges in understanding and implementing security measures, underestimating the risk and consequences of a cyber incident, and a gap in planning for, and responding to, cyber incidents. Further, these barriers can vary among the micro, small and medium enterprises.

SMEs are embracing emerging technologies, with 84% of Australian SMEs having adopted online services and relying on up to 30 separate technologies (Cynch 2021). Most medium firms plan to maintain or increase their investment in the cyber fitness of their business. Australian SMEs recognise the need to look beyond technical solutions when addressing their risks, and 72% of Australian SMEs consider cybersecurity as very important (Cynch 2021). The Australian Information Security Association (AISA) identifies the lack of adequate cybersecurity skills and examines the impact of COVID-19 among micro, small and medium enterprises in New South Wales (NSW). NSW service providers have found it very difficult to attract and retain people with skills in strategy, risk and governance, compliance, and consultancy (AISA 2021). For example, ACSC found that an Australian SME network had been attacked by a malicious cyber adversary and that the SME network had links to national security projects. According to the ACSC analysis, the adversary attacker had access to the SME network for a long period of time and was able to download a vast amount of data. The analysis showed that the adversary was able to gain access to the company's network by misusing an internet-facing server, and then had moved across the entire network with the support of administrative credentials. This enabled the attacker to install multiple web shells in the victims' network; a script was uploaded to a webserver empowering remote administration of the machine, to obtain and retain future access. An investigation revealed that 'sensitive data was freely accessible due to seemingly no encryption or access control measures being in place', so sensitive data was found to be freely accessible due to the apparent absence of encryption or access control measures. The key finding was that the SME did not correctly configure the security features of their systems (ACSC 2018). This highlights that privacy, trust, and security challenges in cloud environments remain a challenge, in particularly for SMEs.

## 3 Research Method

This study employed a descriptive analysis methodology to examine what key cybersecurity threats, vulnerabilities, and security practices adopted in cloud context among Australian micro, small and medium enterprises. Descriptive statistics describes the characteristics of a data set (Amrhein et al. 2019). As the preliminary stage, we analysed these descriptive statistics to get an understanding about the similarities and the differences of cyber practices among SME categories. The unit of analysis for this study is the organisation, in particularly Australian SMEs, which are widely distributed geographically throughout all Australian states and territories. In Australia, an estimated 95.5% of SMEs are connected to an Internet service (ACMA 2010). The survey instrument containing appropriate measures and demographics was designed and thoroughly tested. The survey instrument comprised of two sections: (i) cybersecurity threats and vulnerabilities in the use of cloud computing, (ii) cybersecurity solutions for cloud computing. Accordingly, the target population for this quantitative study consisted of IT related decision makers (IT managers or decision makers in the IT section or other authorities) from different types of SMEs in Australia, who were responsible for and authorised to make policy decisions and technology acquisition for their business. The main study was conducted, where data were collected from decision makers (i.e., owner, IT manager or decision-maker in the IT section) in SMEs. An online survey tool was used to collect data. The survey instrument was developed and designed using Qualtrics survey software. Qualtrics is a widely used online software survey platform. The survey link was shared among a group of selected SMEs (1,410 in total) to align with the convenient sampling technique. From this, 339 responses were received, and 289 valid responses were used for subsequent data analysis.

|  | Frequency | % | Valid % | Cumulative % |
|---|---|---|---|---|
| IT related authorities |  |  |  |  |
| Owner | 97 | 33.6 | 33.6 | 33.6 |
| Chief Executive Officer (CEO) | 21 | 7.3 | 7.3 | 40.9 |
| Chief Information Officer (CIO) | 42 | 14.5 | 14.5 | 55.4 |
| Chief Security Officer (CSO) | 11 | 3.8 | 3.8 | 59.2 |
| IT Manager | 28 | 9.7 | 9.7 | 68.9 |
| IT Executive | 53 | 18.3 | 18.3 | 87.2 |
| Other | 37 | 12.8 | 12.8 | 100.0 |
| Number of employees |  |  |  |  |
| 1-4 (micro) | 166 | 57.4 | 57.4 | 57.4 |
| 5-19 (small) | 71 | 24.6 | 24.6 | 82.0 |
| 20-199 (medium) | 52 | 18.0 | 18.0 | 100.0 |
| State/Territory |  |  |  |  |
| NSW | 83 | 28.7 | 28.7 | 28.7 |
| VIC | 71 | 24.6 | 24.6 | 53.3 |
| QLD | 57 | 19.7 | 19.7 | 73.0 |
| WA | 35 | 12.2 | 12.2 | 85.2 |
| SA | 22 | 7.6 | 7.6 | 92.8 |
| TAS | 11 | 3.8 | 3.8 | 96.6 |
| NT | 3 | 1.0 | 1.0 | 97.6 |
| ACT | 7 | 2.4 | 2.4 | 100.0 |

*Table 1. Demographic characteristics of responding organisations*

The following demographic characteristics of these responses were considered: the title or position of respondents, the organisations size, where the business operated (state or territory), and cloud usage. Table 1 presents the descriptive statistics that were used. The employment profile of the respondents was examined. Most of the respondents (33.6%, n=97) hold the ownership position of organisation. The remaining 67% other decision makers such as chief executive officer (CEO) (7.3%, n=21), chief information officer (CIO) (14.5%, n=42), chief security officer (CSO) (3.8%, n=11), IT manager (9.7, n=28%), IT executive (18.3%, n=53), and other (12.8%, n=37). According to the organisational profile based on total number of employees at the time of this research 57.4% (n=166) of respondents are from organisations within the range of 1 to 4 employees, 24.6% in the range of 5 to 19 employees, and 18% are in the range of 20 to 199 employees. The states with larger populations provided higher response rates. There are 28.7% (n=83) of the respondents who from New South Wales (NSW), Victoria (VIC) (24.6%; n=75), Queensland (QLD) (19.7%; n=57). The remaining organisations came from West Australia (WA) (12.2%), South Australia (SA) (7.6%), Tasmania (TAS) (3.8%), Northern Territory (NT) (1%), and the Australian Capital Territory (ACT) (2.4%). In summary, the analysis of the demographic characteristics of the surveyed SMEs provides insight into the general profile of such SMEs. The surveyed SMEs are also in diverse industries with respect to a range of sizes and the industry type.

## 4 Data Analyses and Results

### 4.1 Cloud Security Threats and Vulnerabilities

The objective of this study was to understand what security issues are related to cloud computing face by three categories within the SME classification. Cross tabulation analyses were conducted by examining relationships between security threats and vulnerabilities by size of enterprise (Table 2). It consisted of several areas: (i) unauthorised access, (ii) insecure interfaces (APIs), (iii) external sharing of data, (iv) hijacking of accounts, services, or traffic, (v) malicious insiders, (vi) malware/ phishing/ ransomware, (vii) misconfiguration of the cloud platform, and (viii) denial of service attacks and other not specified in the list.

|  | Micro | Small | Medium | $X^2$ |
|---|---|---|---|---|
| Unauthorised access | 21.9% (68) | 19.8% (41) | 20.6% (27) | 1.54 |
| Insecure interfaces (APIs) | 15.8% (49) | 11.1% (23) | 9.9% (13) | 2.43 |
| External sharing of data | 18.0% (56) | 14.5% (30) | 13.7% (18) | 1.71 |
| Hijacking of accounts, services, or traffic | 10.0% (31) | 13.5% (28) | 9.2% (12) | 2.15 |
| Malicious insiders | 5.1% (16) | 7.2% (15) | 13.0% (17) | 6.96* |
| Malware/Phishing/Ransomware | 13.8% (43) | 10.6% (22) | 9.9% (13) | 1.86 |
| Misconfiguration of the cloud platform | 6.8% (21) | 11.1% (23) | 7.6% (10) | 3.08 |
| Denial of service attacks | 3.2% (10) | 6.8% (14) | 6.9% (9) | 3.96 |
| Other | 5.5% (17) | 5.3% (11) | 9.2% (12) | 3.47 |

Note. Percentages indicate the percentage of security threats and vulnerabilities and size of enterprise. Statistics indicate whether percentages significantly vary across enterprises size for each of the category, * $p < 0.05$.

*Table 2. Cloud security threats and vulnerabilities*

The data were analysed and compared to identify significant differences of threats and vulnerabilities, and the results were generated into tabular format. These threats and vulnerabilities accounted for a high proportion in the micro sized enterprises (unauthorised access, 21.9%; external sharing of data, 18%; insecure interfaces, 14.5%; malware/phishing/ransomware, 13.8%) and the small (unauthorised access, 19.8%; external sharing of data, 15.3%; hijacking of accounts services or traffic, 14.3%; insecure interfaces, 11.1%; misconfiguration of the cloud platform, 11.1%; malware/phishing/ ransomware, 10.6%), but some threats and vulnerabilities are significant drop in the medium size organisations (unauthorised access, 20.6%; external sharing of data, 13.7%; malicious insiders, 13%). Medium sized enterprises are more than twice as likely to have a malicious insider attack when compared to a micro business. But there are no differences between micro and small, nor between small and medium sized businesses ($x^2 = 6.96$, $p < 0.05$). All other threats and vulnerabilities share the almost same pattern across sized of organisations.

### 4.2 Cloud Security Practices

In order to understand how SMEs face cloud security practices for different types of SMEs, cross tabulation analyses were conducted by examining relationships between security solutions by type of enterprise. It consisted of several areas: (i) data encryption, (ii) encrypted cloud service, (iii) security information event management, (iv) trained Cloud security professional, (v) vulnerability assessment, (vi) access control, (vii) log management and analytics, (viii) privileged access management, (ix) data leakage prevention and (x) security insurance and other not specified in the list. The data were analysed and compared to identify significant differences of security solutions, and the results were generated in Table 3.

|  | Micro | Small | Medium | $X^2$ |
|---|---|---|---|---|
| Data encryption | 11.1% (35) | 14.9% (39) | 18.1% (44) | 7.26* |
| Encrypted cloud service | 18.4% (58) | 15.7% (41) | 14.0% (34) | 3.89 |
| Security Information Event Management | 7.6% (24) | 10.7% (28) | 12.3% (30) | 5.08* |
| Trained cloud security professional | 7.3% (23) | 12.6% (33) | 13.2% (32) | 8.32** |
| Vulnerability assessment | 4.4% (14) | 8.0% (21) | 7.4% (18) | 4.97* |
| Access control (e.g., cloud access security brokers) | 13.7% (43) | 10.7% (28) | 8.6% (21) | 1.81 |
| Log management and analytics | 3.5% (11) | 3.1% (8) | 4.9% (12) | 2.01 |
| Privileged access management (PAM) | 9.8% (31) | 5.4% (14) | 6.2% (15) | 2.75 |
| Data leakage prevention | 10.8% (34) | 6.1% (16) | 4.9% (12) | 5.02* |
| Security Insurance | 6.0% (19) | 6.9% (18) | 5.8% (14) | 0.74 |
| Other | 7.3% (23) | 5.7% (15) | 4.5% (11) | 3.53 |

Note. Percentages indicate the percentage of security practices and size of enterprises. Statistics indicate whether percentages significantly vary across enterprises size for each of the category, * $p < 0.05$; ** $p < 0.01$.

*Table 3. Cloud security practices*

SMEs are considered to have better protection against cloud computing threats if they are using encrypted cloud service and data encryption. Cloud security solutions accounted for a lower proportion in micro sized organisations (encrypted cloud service, 18.4%; access control, 13.7%; security insurance, 11.1%) and the small (encrypted cloud service, 15.7%; data encryption, 14.9%; trained cloud security professional, 12.6%; security information event management, 10.7%; access control, 10.7%). Data encryption and trained cloud security professionals were found in a significantly higher percentages of medium and small sized organisations (data encryption, 18.1%; encrypted cloud service, 14%; trained cloud security professional, 13.2%; security information event management, 12.3%). Medium sized organisations reveal a higher percentage of responses for data encryption, trained Cloud security professionals, and security information event management ($x^2 = 7.26$, $p < 0.05$; $x^2 = 5.08$, $p < 0.05$; $x^2 = 8.32$, $p < 0.01$, respectively) and small-sized organisations for vulnerability assessment ($x^2 = 4.97$, $p < 0.05$) which were significant across the size of the enterprises. In addition, security solutions indicating the use of data leakage prevention technologies were higher for micro sized businesses when compared to both small and medium sized businesses ($x^2 = 5.02$, $p < 0.05$).

## 5  Discussion

Cloud computing is a combination of many existing and emerging technologies like the Internet, networking, operating systems, hardware, software, middleware, virtualisation, and multi-tenancy (Tsai et a. 2010). When cloud computing integrates all of the above technologies, the existing challenges and issues become even more challenging and demanding. If the data stored in cloud computing is sensitive, then cybersecurity becomes an utmost important aspect. The cybersecurity threats and vulnerabilities related to cloud computing that affect SMEs, may differ from large enterprises because of unique contexts. This study found that the major cybersecurity issues in the context of cloud computing that are related to SMEs, include unauthorised access, external sharing of data, malware, phishing, ransomware, misconfiguration of the cloud platform, malicious insiders and insecure interfaces being exploited, among others. The risks of data loss and untrusted server manipulation are key issues faced by SMEs operating in the cloud environment, and such instances threaten the data integrity of organisations. Malicious insiders, malware, phishing, and ransomware present significant threats for SMEs in relation to the availability of information and information systems in the cloud, as well as the cost of using the technology.

Cybersecurity issues include a lack of adequate security standards to ensure robust security for enterprises, trust issues when working over the Internet, and compliance issues. For the cloud infrastructure, the major security issues for SMEs include insecure API interfaces, security misconfiguration, server location and data backup, as well as the multi-tenancy characteristics of the cloud environment. The study found that the major cybersecurity threats and vulnerabilities related to cloud computing does not depend on the size of the businesses, with the exception of threats from malicious insiders. Malicious insiders in the cloud may be current or former employees, contractors, or a trusted service provider from the cloud who misuses their authority to access critical assets in the cloud without permission from the organisation. Insider threats are a major security issue for any organisation. A malicious insider already has authorised access to an organisation's network and some of the sensitive resources that it contains. Medium sized firms can have many third-party vendors with granted access to the organisation's internal networks, making the network even more vulnerable to security breaches compared to micro and small businesses. In addition, illegal practices related to IT are much easier to log, notice, and correct on a smaller network than on a network with many employees.

This study also examined the current cybersecurity practices, techniques, and challenges against security issues in the cloud computing context. This study found that the most common security methods used by enterprises include data encryption, security information event management, trained cloud security professionals, vulnerability assessment, access control, privileged access management as well as data leakage prevention. However, there was no significant difference for encrypted cloud service, access control, privileged access management, log management and analytics and security insurance between micro, small and medium sized enterprises. This study reveals that in terms of security practices and techniques such data encryption, security information event management, trained cloud security professionals, vulnerability assessment and data leakage prevention are all experienced significantly differently between medium, small, and micro enterprises. SMEs reported better security practices and techniques on data encryption, security information event management, trained cloud security professionals, and vulnerability assessment than the micro firms. However, micro industries practiced better data leakage prevention when compared to the other two larger sized industries. This suggests that micro and small firms are less affected by this aspect of security when compared to medium sized enterprises. Micro enterprises often lack dedicated cybersecurity resources, lack funds, time and expertise to protect against cyber-threats. This makes them especially vulnerable to highly adaptable and cunning cybercriminals who look to exploit vulnerable micro businesses. Micro and small enterprises in Australia have a somewhat weak understanding of cybersecurity, control methods and security technologies in a cloud context, and they typically depend on third parties.

# 6   Conclusion

This study reveals that Australian SMEs are vulnerable to cyber-attacks and further, they do not follow good cybersecurity practices. The study findings have both theoretical and practical contributions, especially with constant technological advances and emerging cybersecurity threats. Security threats and vulnerabilities, as well as opportunities, apply to many SMEs, so this investigative work should be continued. The study adds to the growing body of literature on security threats and vulnerabilities to enterprises in cloud environments by examining the priority threats unique to all sizes of small, medium and micro sized enterprises. This study found significant differences in cybersecurity practices among micro, small, and medium enterprises. For example, data encryption, security information event management, trained cloud security professionals, vulnerability assessment, access control, privileged access management and data leakage prevention all exhibit lower levels of impact for micro and small sized firms, when compared to medium sized firms. However, and interestingly, micro sized enterprises exhibit higher levels of data leakage prevention compared to small and medium sized enterprises.

The main contribution of this paper was highlighting the diversity of cybersecurity issues and practices by dividing the sample into three segments namely micro, small, and medium. The study bridges the research gap and provides insightful evidence on these issues, especially within the context of Australian businesses of this size. Results indicate that micro, small and medium sized enterprises face similar kinds of cybersecurity threats and vulnerabilities, but significant differences in their cloud security practices are revealed. A possible reason for this might be due to resource limitations, or perhaps they may face inadequate expertise regarding cloud computing and its practices. Also, generally SMEs are less concerned about security and privacy threats, in part because they do not have dedicated IT staff and the associated knowledge. This study has several practical contributions. The new understandings (i.e., that security and privacy factors do not impose significant influences on SMEs) enable SMEs, cloud service providers, IT practitioners and policy makers to concentrate on other critical factors that have more impact on cloud computing adoption, such as broadband affordability and speed. As SMEs are generally not capable of significant investment on IT, it is essential for cloud service providers, professional bodies, and the government to devise strategies for the widespread adoption of cloud computing for these businesses. For cloud service providers it is important to identify the differences among different SMEs and provide purpose driven, bespoke cybersecurity solutions.

This research focuses only on SMEs within Australia and should be considered carefully for other regions. For example, results may not be directly applicable to SMEs from other parts of the world. Hence, this study would have benefited from the inclusion of perspectives of SMEs from other countries. Further, this study is limited in terms of its quantitative data. Qualitative research, such as case studies and longitudinal studies, are required to gain a more meaningful understanding of this phenomenon. Future research could focus on case studies to explore security and privacy impacts among SMEs. Furthermore, future studies could focus on mitigation strategies for security and privacy issues.

## Acknowledgements

This research was supported by the RMIT international tuition fee scholarship, the RMIT University and the Centre for Cyber Security Research and Innovation (CCSRI).

## Copyright